\documentclass[aps,preprint]{revtex4}
\usepackage{amsmath}
\usepackage{bm}
\usepackage{graphicx}

\newcommand{\va}{\mathbf{a}}

\newcommand{\vvr}{\mathbf{r}}

\newcommand{\vcr}{\mathbf{R}}

\newcommand{\hb}{\mathbf{\hat b}}

\newcommand{\hr}{\mathbf{\hat r}}

\newcommand{\hcr}{\mathbf{\hat R}}

\begin{document}

\title{Hyperspherical harmonics with arbitrary arguments}

\author{A.~V.~Meremianin}
\email{meremianin@phys.vsu.ru}
\affiliation{Department of Theoretical Physics, 
Voronezh State University, 394006, Voronezh,  Russia}

\date{\today}

\begin{abstract}
The derivation scheme for hyperspherical harmonics (HSH) with arbitrary
arguments is proposed.
It is demonstrated that HSH can be presented as the product of
HSH corresponding to spaces with lower dimensionality multiplied by the
orthogonal (Jacobi or Gegenbauer) polynomial.
The relation of HSH to quantum few-body problems is discussed.
The explicit expressions for orthonormal HSH in spaces with dimensions from
$2$ to $6$ are given.
The important particular cases of four- and six-dimensional spaces are
analyzed in detail and explicit expressions for HSH are given for several
choices of hyperangles.
In the six-dimensional space, HSH representing the kinetic energy operator
corresponding to i) the three-body problem in physical space and ii)
four-body planar problem are derived.
\end{abstract}


\maketitle

\section{Introduction}
\label{sec:introduction}

The hyperspherical harmonics (HSH) are an important tool in the study of
quantum few-body systems.
This is caused by the fact that the kinetic-energy operator of an $N$-particle
system is equivalent to the Laplace operator in the space of $3N$ dimensions
(or $2N$, in the case of planar systems).
The amount of papers in which HSH are applied to specific physical problems is
very large.
We mention only few recent review articles \cite{lin95:_hypersph_prep,%
aquilanti01:_rev_hypsp_mom_space,Kuppermann94,fedorov01:_3bd_short_range} and
a book \cite{avery00:_hspher_book}.

HSH are functions of $d-1$ dimensionless variables (hyperangles) which
describe the points on the hypersphere.
Of course, the choice of hyperangles is not unique and it is the matter of
convenience.
The only exception is the three-dimensional space, where the arguments of the
spherical harmonics are conventional spherical angles 
$\theta=\arccos z/\sqrt{x^2+y^2+z^2}$ and $\phi=\arctan x/y$.
The generalization of this definition on spaces with $d > 3$ is the following.
Let $\vvr=(x_1, x_2, \ldots , x_d)$ be the $d$-dimensional radius-vector, then
the hyperspherical angles are defined by the equations \cite{Bateman-II},
\begin{equation}
  \label{eq:l-3}
  \begin{split}
 x_1 &= R \, \cos \theta_1, \\
 x_2 &= R \, \sin \theta_1 \cos \theta_2, \\
 x_3 &= R \, \sin \theta_1 \sin \theta_2 \cos\theta_3, \\
 & \ldots \\
 x_{d-1} &= R\, \sin \theta_1 \ldots \sin\theta_{d-2}\, \cos\phi\\
 x_{d} &= R\, \sin \theta_1 \ldots \sin\theta_{d-2}\, \sin\phi\\
  \end{split}
\end{equation}
where $\phi \in [0,2\pi)$, $\theta_{\kappa} \in [0, \pi]$, 
$\kappa=1,2,\ldots, d-2$.

The $J$-th rank HSH $H^J_\epsilon(\hr)$ is defined by the Laplace equation
\begin{equation}
  \label{eq:laplace}
  \Delta r^J H^J_\epsilon(\hr)=0,
\end{equation}
where $\Delta = \sum_{k=1}^d \partial^2 /\partial x_k^2$ and $\epsilon$
denotes the set of indices which label different HSH of the same rank.

The explicit expression for HSH depending on the hyperspherical angles
(\ref{eq:l-3}) is well-known (see eq.(11.2.23) of
\cite{Bateman-II}).
It is given by the product of Gegenbauer (ultraspherical) polynomials, so that
$k$-th polynomial in the product depends on $\cos \theta_k$.

\begin{figure}[ptbh]
\centering
\includegraphics[width=5cm]{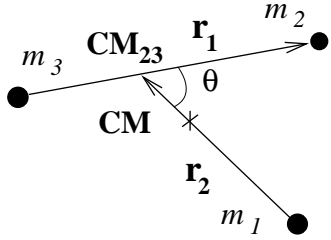}
\caption{Jacobi vectors \cite{smirnov_shitikova77:_k_harm} for the three-body
  system. $CM_{23}$ is the center of mass of the pair $m_{2}$,$m_{3}$.}%
\label{fig:3bd-jacobi}%
\end{figure}

However, in many physical applications the set of hyperspherical angles is not
convenient.
For example, in the quantum three-body problem one has to deal with
six-dimensional vector space being the product of two three-dimensional
spaces corresponding to the Jacobi vectors, see fig.~\ref{fig:3bd-jacobi}.
Thus, in this case it is more adequate to parametrize the arguments of HSH by
the spherical angles of the three-dimensional vectors $\vvr_1,\vvr_2$.
Further aspects of the choice of coordinates in the three-body
problem including analysis of the permutational symmetry are discussed in
\cite{aquilanti:86_hh_3bd}.

HSH in $3N$-dimensional space depending on the spherical angles of $N$
three-dimensional vectors may serve as a convenient basis for the
decomposition of the wave function of the quantum many-particle system.
The arguments of corresponding HSH comprise apart of $2N$ three-dimensional
spherical angles of vectors  $\vvr_1, \ldots \vvr_N$ also $N-1$ hyperangles
which describe the lengths $r_1,r_2,\ldots r_N$.
Again, there is freedom in choice of the hyperangles. 
For example, one can define the hyperangles similarly to (\ref{eq:l-3})
by replacing $x_k \to r_k$.
Such set of the hyperangles was used in 
\cite{knirk74:_HSH_individual_ang_mom} where the explicit expressions for the
corresponding HSH have been derived by means of the explicit solution of the
Laplace equation.

The physical nature of the problem may suggest different choice of the
connection of the hyperangles with the vectors desribing the system.
Thus, the problem is how to derive the explicit expression for HSH
corresponding to various choices of their arguments.
The conventional approach
\cite{Bateman-II,avery85jmp:_hyp_spher,knirk74:_HSH_individual_ang_mom}, 
consists in the transformation of the Laplacian to the desired hyperangular
variables with the subsequent solution of the ensuing partial differential
equation.
Such approach has been used in \cite{aquilanti:86_hh_3bd} where the number of
relations for HSH has been presented using the so-called ``method of trees''.

It is desirable to develop the technique of the derivation of HSH which would
not require the transformation of the Laplacian.
In the presented paper such the technique is proposed.
It allows one to derive the expressions for HSH depending on arbitrary set of
hyperangles.
The technique is based on the concept of zero-length vectors proposed
initially by Cartan \cite{cartan66:_spinors_book,biedenharn}.
The developed calculation scheme is recursive, i.e. HSH in space with higher
dimensions is presented as the product of lower-dimensional HSH with some
weight function being an orthogonal polynomial.
As the examples of the proposed technique, various representations for HSH in
spaces with $d=2,3,4,5,6$ are derived.


\section{The general formalism}
\label{sec:idea}

Below the expression for the scalar product of HSH is derived in
sec.~\ref{sec:scalar-product-hsh}.
Next, in sec.~\ref{sec:d-dim-hsh} several general expressions for HSH are
derived in terms of lower-rank HSH.
The final expressions for normalized HSH and surface elements in
$d$-dimensional space are given in sec.~\ref{sec:orth-norm}.

\subsection{The scalar product of HSH}
\label{sec:scalar-product-hsh}

It is important that the expression for the scalar product of HSH
can be derived without the knowledge of their explicit form.
In order to demonstrate this, let us consider the scalar function 
$f_J(\vvr,\vvr')$ defined by
\begin{equation}
  \label{eq:gf-0}
  f_J (\vvr, \vvr') = r^{2J+d-2} \, (\vvr' \cdot \nabla)^J \frac{1}{r^{d-2}},
\end{equation}
where $\vvr$, $\vvr'$ are $d$-dimensional vectors.
Obviously, $f_J$ is an homogeneous polynomial of degree $J$ with respect to
the components of $\vvr$ or $\vvr'$.
Namely,
\begin{equation}
  \label{eq:gf-00}
  f_J (\alpha \vvr, \beta \vvr') = (\alpha \beta)^J\,
f_J (\vvr, \vvr'), \quad \alpha,\beta = \mathrm{const}.
\end{equation}
One can prove that $f_J$ satisfies the Laplace equation with respect to both
vectors $\vvr$ and $\vvr'$.
Indeed,
\begin{equation}
  \label{eq:gf-0a}
  \Delta' f_J = 
J\,r^{2J+d-2} \,
 (\nabla \cdot \nabla')\, (\vvr' \cdot \nabla)^{J-1} \frac{1}{r^{d-2}}
=J(J-1)\, r^{2J+d-2} \,
\Delta \, (\vvr' \cdot \nabla)^{J-2} \frac{1}{r^{d-2}}=0.
\end{equation}
Here, $\Delta'$ and $\nabla'$ act on the components of $\vvr'$.
We have also used the identity $\Delta r^{2-d}=0$ and the fact that
the differential operations are commutative.
The proof of $\Delta f_J =0$ is somewhat more lengthy,
\begin{equation}
  \label{eq:gf-0b}
  \Delta f_J =  2 (2J+d-2) \left( \frac{(J+d-2)}{r^2}\, f_J
+ r^{2J+d-3}\, (\vvr \cdot \nabla) (\vvr' \cdot \nabla)^J
\frac{1}{r^{d-2}}
\right).
\end{equation}
Using the operator identity
\begin{equation}
  \label{eq:gf-0ba}
(\vvr' \cdot \nabla)^J (\vvr \cdot \nabla)
= J\, (\vvr' \cdot \nabla)^J
+ (\vvr \cdot \nabla) (\vvr' \cdot \nabla)^J
\end{equation}
one can transform the second term in~(\ref{eq:gf-0b}) to 
\begin{equation}
  \label{eq:gf-0bb}
 (\vvr \cdot \nabla) (\vvr' \cdot \nabla)^J \frac{1}{r^{d-2}}
= \left[
(\vvr' \cdot \nabla)^J (\vvr \cdot \nabla)- J\, (\vvr' \cdot \nabla)^J
\right] \frac{1}{r^{d-2}}
= - (J+d-2) (\vvr' \cdot \nabla)^J \frac{1}{r^{d-2}}.
\end{equation}
The substitution of this identity into (\ref{eq:gf-0b}) completes the proof of
the equation $\Delta f_J =0$.

The function $f_J (\vvr, \vvr')$ is $J$-th order homogeneous polynomial
satisfying the Laplace equations $\Delta f_J = \Delta' f_J =0$ is
\textit{scalar} and, therefore, it can be nothing else than the scalar product
of two HSH of rank $J$,
\begin{equation}
  \label{eq:gf-s1}
  f_J (\vvr, \vvr') = (r r')^J \sum_{\epsilon} 
\left[ H^J_\epsilon (\hr') \right]^*
 H^J_\epsilon (\hr).
\end{equation}
The explicit form of $f_J$ can be calculated as follows.
First, we note that the gradient operators in (\ref{eq:gf-0}) can be replaced
with the derivatives,
\begin{equation}
  \label{eq:gf-s2}
  f_J = r^{2J+d-2} \frac{d^J}{d t^J} 
\frac{1}{|\vvr + t\, \vvr'|^{d-2}} \biggr|_{t=0}
= \frac{d^J}{d t^J} 
\frac{(-1)^J \, r^{2J}}%
{( 1 - 2 \xi (t r'/r) + (t r'/r)^2 )^{d/2-1}} \biggr|_{t=0},
\end{equation}
where $\xi = (\hr' \cdot \hr)$ is the cosine of the angle between
$\vvr'$ and $\vvr$.
Noting that the function to be differentiated coincides with the generating
function for Gegenbauer (or ultraspherical) polynomials \cite{Bateman-II} we
can write,
\begin{equation}
  \label{eq:gf-s3}
  f_J = (-1)^J\, r^{2J} \frac{d^J}{d t^J} 
\sum_{n=0}^\infty \left( \frac{t r'}{r} \right)^n\,
C^{d/2-1}_n (\xi) \biggr|_{t=0}
= (-r r')^J \, J!\, C^{d/2-1}_J (\xi).
\end{equation}
At this stage, we note that HSH are defined up the some normalization factor
which can be an arbitrary number independent of $r,r'$.
Thus, the scalar product of HSH (\ref{eq:gf-s1}) expresses as
\begin{equation}
  \label{eq:gf-s4}
   \sum_{\epsilon} 
\left[ H^J_\epsilon (\hr') \right]^* H^J_\epsilon (\hr)
= A^{(d)}_J \, C^{d/2-1}_J (\hr' \cdot \hr),
\end{equation}
where $A^{(d)}_J$ is the normalization constant and 
$C^{d/2-1}_J(\hr' \cdot \hr)$ is the $J$-th order Gegenbauer polynomial.

\subsection{The construction of $d$-dimensional HSH}
\label{sec:d-dim-hsh}

The main idea is based on the fact that the scalar product 
$(\va \cdot \vvr)^J$ for ``zero-length'' vectors
$\va$ (i.e. for $(\va\cdot\va)=0$) satisfies the Laplace equation. 
The proof is very simple,
\begin{equation}
  \label{eq:gf-1}
  \Delta (\va \cdot \vvr)^J = 
J \, (\va \cdot \nabla)\, (\va \cdot \vvr)^{J-1}
= J(J-1)\, (\va \cdot \va)\, (\va \cdot \vvr)^{J-2} =0.
\end{equation}
The condition $(\va\cdot\va)=0$ means that some components of the
zero-length vector $\va$ must be complex numbers.
Let us choose the components of $\va$ to be
\begin{equation}
  \label{eq:gf-2}
  \va = (\hb_\kappa, i\, \hb_{d-\kappa}), \quad \kappa < d,
\end{equation}
where $d$ is the dimensionality of $\va$ and $\vvr$.
The \textit{real} unit vectors $\hb_\kappa$ and $\hb_{d-\kappa}$ have
dimensionalities $\kappa$ and $d-\kappa$, respectively.
It is easy to see that the zero-length condition is met
\begin{equation}
  \label{eq:gf-3}
  \va \cdot \va 
= (\hb_\kappa)^2 + (i\, \hb_{d-\kappa})^2 = 1-1=0.
\end{equation}
Let us parametrize the $d$-dimensional radius-vector $\vvr$ as
\begin{equation}
  \label{eq:gf-3a}
  \vvr = (\vvr_\kappa, \, \vvr_{d-\kappa}).
\end{equation}
It is more convenient to work with unit vectors so that
\begin{equation}
  \label{eq:gf-3b}
  \begin{split}
  \vvr_\kappa =& r_\kappa \, \hr_\kappa = r \, \cos \theta_\kappa \, 
    \hr_\kappa, \\
  \vvr_{d-\kappa} =& r_{d-\kappa} \, \hr_{d-\kappa} = r \, \sin \theta_\kappa\,
   \hr_{d-\kappa},
  \end{split}
\end{equation}
where $\theta_\kappa \in [0, \pi/2]$ and the hyper-radius is defined by 
$r^2=r_\kappa^2+r_{d-\kappa}^2=\sum_{n=1}^d x_n^2$.
Note that the above parametrization implies that both $\kappa >1$ and 
$(d-\kappa) >1$.
It is important to note that we are not making any assumptions about the
parametrization of the unit vectors $\hr_\kappa$ and $\hr_{d-\kappa}$.

Suppose that explicit expressions for HSH with the dimensionalities $\kappa$,
$(d-\kappa)$ are known.
Then, one can obtain an expression for the $d$-dimensional HSH as the product
of known HSH with some weight function.
We begin the proof of this statement by writing
\begin{equation}
  \label{eq:gf-4}
  (\va \cdot \vvr)^J
=[(\hb_{\kappa} \cdot \vvr_{\kappa}) 
+ i\, (\hb_{d-\kappa} \cdot \vvr_{d-\kappa})]^J
=\sum_{q=0}^J \binom{J}{q}
i^{J-q}\, (\hb_{\kappa} \cdot \vvr_{\kappa})^q \,
(\hb_{d-\kappa} \cdot \vvr_{d-\kappa})^{J-q}.
\end{equation}
Now we expand the scalar products over the Gegenbauer polynomials which, in
turn, are the scalar products of HSH.
Namely,
\begin{equation}
  \label{eq:gf-5}
  (\hb_{\kappa} \cdot \vvr_{\kappa})^q
= r_\kappa^q \sum_{l=0}^q B^{(\kappa)}_{ql}\,
C^{\kappa/2-1}_{l} (\hb_\kappa \cdot \hr_\kappa)
= r_\kappa^q \sum_{l=0}^q \frac{B^{(\kappa)}_{ql}}{A^{(\kappa)}_l} \, 
\sum_{\epsilon} \left[ H^l_{\epsilon} (\hb_\kappa) \right]^* 
H^l_{\epsilon} (\hr_\kappa).
\end{equation}
Here, $\epsilon$ denotes the set of $(\kappa-2)$ projection indices of HSH, 
$B^{(\kappa)}_{ql}$ are the expansion coefficients which can be calculated
using the orthogonality of Gegenbauer polynomials.
Omitting details of the computations, we present only the final expression for
$B^{(\kappa}_{ql}$,
\begin{equation}
  \label{eq:gf-5a}
B^{(\kappa)}_{ql} = \sigma_{q,l} \,
\frac{q!\, (l+\kappa/2-1)\,\Gamma(\kappa/2-1)\, \Gamma(n+1/2) }
{\sqrt \pi \, 2^l\, (q-l)!\, \Gamma(l+\kappa/2+n)}, \quad
q-l=2n,
\end{equation}
where integer $n \ge 0$ and $\sigma_{q,l}=[1+(-1)^{q-l}]/2$.
Thus, $B^{(\kappa)}_{ql}=0$ if $q$ and $l$ have different parities.

Substituting equation (\ref{eq:gf-5}) and the similar equation for 
$(\hb_{d-\kappa} \cdot \vvr_{d-\kappa})^{J-q}$ into (\ref{eq:gf-4}), the
scalar product $(\va \cdot \vvr)^J$ can be written as
\begin{equation}
  \label{eq:gf-6}
 (\va \cdot \vvr)^J = r^J
 \sum_{l,l'} \left[ H^l_{\epsilon} (\hb_\kappa) 
 H^{l'}_{\epsilon'} (\hb_{d-\kappa})  \right]^* \,
\frac{H^J_{l \epsilon,\, l' \epsilon'} (\hr)}%
{A^{(\kappa)}_l\, A^{(d-\kappa)}_{l'}},
\end{equation}
where the functions $H^J_{l \epsilon,\, l' \epsilon'} (\hr)$ are defined by
the product
\begin{equation}
  \label{eq:gf-7}
H^J_{l \epsilon,\, l' \epsilon'} (\hr) =
H^l_{\epsilon} (\hr_\kappa)  H^{l'}_{\epsilon'} (\hr_{d-\kappa})\,
h^J_{ll'} (\theta_\kappa),
\end{equation}
where $h^J_{ll'} (\theta_\kappa)$ denotes the summation
\begin{equation}
  \label{eq:gf-8}
  h^J_{ll'} (\theta_\kappa) = 
\sum_{q} \binom{J}{q} i^{J-q}\, 
(\cos \theta_\kappa)^q\, (\sin \theta_{\kappa})^{J-q}\,
B^{(\kappa)}_{ql}\, B^{(d-\kappa)}_{(J-q) \, l'}.
\end{equation}
Noting that the action of the Laplace operator on (\ref{eq:gf-6}) gives
$\Delta (\va \cdot \vvr)^J =0$, and, since $\hb_\kappa$ and $\hb_{d-\kappa}$
are arbitrary vectors, we arrive at the identity
\begin{equation}
  \label{eq:gf-9}
  \Delta \left[ r^J  H^J_{l \epsilon,\, l'\epsilon'} (\hr) \right] =0.
\end{equation}
Therefore, $H^J_{l \epsilon,\, l'\epsilon'}$ are $d$-dimensional hyperspherical
harmonics of the rank $J$ with projection indices $l\epsilon,\, l'\epsilon'$.

The explicit form of the functions $h^J_{ll'} (\theta_\kappa)$ defined by
(\ref{eq:gf-8}) is derived in the Appendix \ref{sec:app-hj}, where it 
is shown that $h^J_{ll'}$ are proportional to the Jacobi polynomials, see
eq.~(\ref{eq:app-2}).

The above consideration was performed under an assumption that $\kappa>1$
and $d-\kappa>1$.
Thus, the case $\kappa=1$ needs special consideration.
First, we parametrize the radius-vector $\vvr$ as
\begin{equation}
  \label{eq:k1-1}
  \vvr =
r \, (\cos \theta, \, \sin \theta\, \hr_{d-1} ), \quad
\theta \in [0, \pi].
\end{equation}
Again, no assumptions are made about the parametrization of the components of
the unit $(d-1)$-dimensional vector $\hr_{d-1}$.

We choose the parametrization of the zero-length vector $\va$ to be
\begin{equation}
  \label{eq:k1-2}
  \va = (1, \, i\, \hb),
\end{equation}
where $\hb$ is the unit real $(d-1)$-dimensional vector.
The expansion of the scalar product $(\va \cdot \vvr)^J$ has the form
\begin{equation}
  \label{eq:k1-3}
    (\va \cdot \vvr)^J
= r^J\, [1 + i\, (\hb \cdot \vvr_{d-1})]^J
=\sum_{q=0}^J \binom{J}{q}
i^{J-q}\, (\cos \theta)^{J-q}\, (\sin \theta)^{J-q}\,
(\hb \cdot \vvr_{d-1})^q.
\end{equation}
Decomposing the scalar product $(\hb \cdot \vvr_{d-1})^q$ using
eq.~(\ref{eq:gf-5}) (where $\kappa=d-1$) we can re-write the above equation as
\begin{equation}
  \label{eq:k1-4}
   (\va \cdot \vvr)^J = r^J
 \sum_{l} \left[ H^l_{\epsilon} (\hb) \right]^* \,
\frac{H^J_{l \epsilon} (\hr)}{A^{(d-\kappa)}_{l}},
\end{equation}
where the $d$-dimensional HSH $H^J_{l \epsilon} (\hr)$ is the
product of $(d-1)$-dimensional HSH and the function of $\theta$
\begin{equation}
  \label{eq:k1-5}
  H^J_{l \epsilon} (\hr)
= g^J_{l} (\theta)\, H^l_{\epsilon} (\hr_{d-1}),
\end{equation}
where $g^J_{l} (\theta)$ are defined similarly to (\ref{eq:gf-8}),
\begin{equation}
  \label{eq:k1-6}
  g^J_{l} (\theta) =
\sum_{q} \binom{J}{q} i^{J-q}\, (\cos \theta)^q\, (\sin \theta)^{J-q}\,
B^{(d-1)}_{ql}.
\end{equation}
This summation evaluates in a closed form as the Gegenbauer polynomial,
see eq.~(\ref{eq:app-g-2}) of Appendix~\ref{sec:app-hj}.


\subsection{Orthogonality and normalization}
\label{sec:orth-norm}

It is important that HSH defined by (\ref{eq:gf-7}) form an orthogonal set on
the $d$-dimensional hypersphere.
In order to demonstrate this we have to derive the expression for the surface
element on the hypersphere
\begin{equation}
  \label{eq:gf-10}
  d \vvr = r^{d-1}\, d r\, d \Omega_d
= d \vvr_\kappa\, d \vvr_{d-\kappa}
= r_\kappa^{\kappa-1} d r_\kappa\, r_{d-\kappa}^{d-\kappa-1} d r_{d-\kappa}
d \Omega_{\kappa}\, d \Omega_{d-\kappa}.
\end{equation}
Noting that $d r_\kappa\, d r_{d-\kappa}= r d r\, d \theta_k$ and using the
hyperspherical parametrization (\ref{eq:gf-3b}) of $r_\kappa,r_{d-\kappa}$,
the surface element of the $d$-dimensional hypersphere is
\begin{equation}
  \label{eq:gf-11}
  d \Omega_d = (\cos \theta_\kappa)^{\kappa-1}\,
(\sin \theta_\kappa)^{d-\kappa-1}\, d \theta_\kappa\, 
d \Omega_{\kappa}\, d \Omega_{d-\kappa}.
\end{equation}
The orthogonality of HSH (\ref{eq:gf-7}) follows from the orthogonality of
$\kappa$- and $(d-\kappa)$-dimensional HSH and properties of Jacobi
polynomials \cite{Bateman-II}.
The same is also true for HSH defined by (\ref{eq:k1-5}).

Assuming that $\kappa$- and $(d-\kappa)$-dimensional HSH 
$H^l_{\epsilon}(\hr_\kappa)$ and $H^{l'}_{\epsilon'}(\hr_{d-\kappa})$ are
normalized, the normalized $d$-dimensional HSH can be written as
\begin{equation}
  \label{eq:gf-12}
 Y^J_{l \epsilon,\, l' \epsilon'} (\hr) =
H^l_{\epsilon} (\hr_\kappa)  H^{l'}_{\epsilon'} (\hr_{d-\kappa})\,
y^J_{ll'} (\theta_\kappa),
\end{equation}
where the functions $y^J_{ll'}$ are proportional to Jacobi polynomials
$P^{(\alpha,\beta)}_\lambda$,
\begin{equation}
  \label{eq:gf-13}
y^J_{ll'} (\theta_\kappa)
= N^{(d,\kappa)}_{J l l'} \,
(\cos \theta_\kappa)^{l}\, (\sin \theta_{\kappa})^{l'}\,
P_{\lambda}^{\left( l'-1+ \frac{d-\kappa}{2}, \,
 l -1 + \frac{\kappa}{2} \right) } (\cos 2 \theta_\kappa),  
\end{equation}
where $\lambda=(J-l-l')/2$.
Note that $y^J_{ll'}$ is non-zero only for $\lambda$  being an integer
number.
The normalization constant in the above equation is defined by
\begin{equation}
  \label{eq:app-4}
 N^{(d,\kappa)}_{J l l'}
= \left[
\frac{(2J-2+d)\,\lambda!\, \Gamma(\lambda+l+l'+d/2-1) }%
{\Gamma(\lambda+l' + (d-\kappa)/2 )\,
\Gamma (\lambda+l +\kappa/2 ) }
\right]^{1/2}. 
\end{equation}
The orthogonality relation for functions $y^J_{ll'} (\theta_\kappa)$ has the
form
\begin{equation}
  \label{eq:gf-14}
  \int_0^{\pi/2} y^{J}_{ll'} (\theta_\kappa)\,
y^{J'}_{ll'} (\theta_\kappa)\, 
(\cos \theta_\kappa)^{\kappa-1}\, 
(\sin \theta_\kappa)^{d-\kappa-1}\, d \theta_\kappa = \delta_{J,J'}.
\end{equation}

In the case $\kappa=1$ the orthonormal HSH can be written as the product of
$(d-1)$-dimensional normalized HSH  $H^l_\epsilon (\hr_{d-1})$ and
the function $y^J_l (\theta)$,
\begin{equation}
  \label{eq:k2-2}
  \begin{split}
   Y^{J}_{l \epsilon} (\hr) &= y^J_l (\theta) \, H^l_\epsilon (\hr_{d-1}), \\
y^J_l (\theta)  &=  N^{(d)}_{J l}\,
 (\sin \theta)^{l}\,
C_{J-l}^{l + \frac{d-1}{2}} (\cos \theta),   
  \end{split}
\end{equation}
where $C_{J-l}^{l + \frac{d-1}{2}} (\cos \theta)$ is the Gegenbauer polynomial
and the normalization coefficient is given by
\begin{equation}
  \label{eq:k2-4}
   N^{(d)}_{J l}
= \Gamma \left( l+ \frac{d-1}{2} \right)
\left[
\frac{2^{2l+d}\, (J+(d-1)/2)\, (J-l)!  }%
{4 \pi\, (J+l+d-2)!}
\right]^{1/2}.
\end{equation}
The orthogonality relation for the functions $y^J_l$ reads
\begin{equation}
  \label{eq:k2-5}
  \int_0^\pi  y^J_l (\theta)\, y^{J'}_l (\theta)
(\sin \theta)^{d-1}\, d \theta = \delta_{J,J'}.
\end{equation}
The above equations (\ref{eq:gf-12})--(\ref{eq:k2-5}) constitute the main
results of the presented paper.


\section{HSH in spaces with $d=2,\ldots 6$}
\label{sec:hsh-2-6}

Below we consider HSH in spaces with dimensionalities from two to six.
This is necessary in order to establish the connection of the derived HSH with
the expressions existing in literature (if any).

\subsection{Two- and three- dimensional HSH}
\label{sec:23d}

The zero-length vector $\va=(1, i)$ and the radius-vector $\vvr=(x, y)$.
Their scalar product is
\begin{equation}
  \label{eq:2d-1}
  (\va \cdot \vvr)^m = (x + i \, y)^m = r^m\, e^{i\,m\phi},
\end{equation}
where $\phi$ is the polar angle, $\phi \in [0, 2\pi)$.
Thus, the two-dimensional spherical (or, better, circular) normalized
harmonics are 
\begin{equation}
  \label{eq:2d-2}
  Y_m (\phi) = \frac{e^{i\,m \phi}}{\sqrt{2\pi}},
\end{equation}
Note that $m \ge 0$ is the rank of HSH.
However, by considering $\va=(1,-i)$ one obtains that $\exp{(-m\phi)}$ is HSH
too.
Therefore, in (\ref{eq:2d-2}) the index $m$ can be $\pm1, \pm2, \ldots$.
Note also that $Y_{m}(-\phi)=Y_{-m} (\phi) = Y^*_m (\phi)$.

The above expression for the two-dimensional HSH allows one to obtain
three-dimensional HSH by using eq.~(\ref{eq:k2-2}) of
sec.~\ref{sec:orth-norm}.
The radius-vector $\vvr$ we decompose into the direct product of the
one component parameter $\cos \theta$ and two-dimensional vector, so that
\begin{equation}
  \label{eq:3da-0}
  \vvr = r\, (\cos \theta, \, \sin \theta\, \hr_2)
= r \, (\cos \theta,\, \sin \theta \cos \phi, \, 
\sin \theta \sin \phi).
\end{equation}
Inserting $d=2$ and $l=m$ into eqs.~(\ref{eq:k2-2}) and (\ref{eq:k2-4})
we obtain
\begin{equation}
  \label{eq:3da-1}
  Y^J_m (\hr) = Y^J_m (\theta, \phi) = (2m-1)!!\, 
\sqrt{\frac{2J+1}{4\pi}\frac{(J-m)!}{(J+m)!} }\, e^{i\, m \phi}\,
(\sin \theta)^m\,
C^{m+1/2}_{J-m} (\cos \theta).
\end{equation}
These functions differ from the conventional spherical harmonics
$Y_{Jm}(\theta,\phi)$ only by the phase factor $(-1)^m$.
Clearly, all properties of $Y^J_m$ are the same as of spherical harmonics
(see e.g.~\cite{Varsh}) and we will not discuss them further.


\subsection{Four-dimensional HSH}
\label{sec:four-d}

The importance of the four-dimensional HSH stems from the fact that they
represent the wave function of the hydrogen atom in the momentum space
\cite{fock35:_o4}.
Also, in the momentum space HSH can be used as the Sturmian basis set which
was successfully applied to many problems of quantum physics and chemistry
\cite{avery04:_shibuya_wulfman_fock,avery00:_hspher_book,%
aquilanti01:_rev_hypsp_mom_space,meremianin06:_fock_jpa}.

There are two possibilities of representing the $4D$-vector: it can be split
into either $(1+3)$- or $(2+2)$-dimensional vectors.
Below, the explicit expressions for the corresponding HSH are derived.


\subsubsection{The parametrization by $1D+3D$ vectors}
\label{sec:3d-1d}

The radius-vector is $\vcr=(\cos \omega, \sin \omega\, \hr)$, where
$\hr=(\cos \theta, \, \sin\theta\cos\phi,\, \sin\theta \sin\phi)$ is the unit
three-dimensional vector.

According to the equations (\ref{eq:k2-2}) and (\ref{eq:k2-4}), we have
\begin{equation}
  \label{eq:4da-1}
  Y^J_{lm} (\hcr) = Y^J_{lm} (\omega, \theta, \phi)
= l!\,
\sqrt\frac{2(J+1)\, (J-l)!}{ \pi \, (J+l+1)!}
(2 \sin \omega)^l\,C^{l+1}_{J-l} (\cos\omega)\,
Y_{lm} (\theta,\phi).
\end{equation}
The orthogonality relation for these harmonics has the form
\begin{equation}
  \label{eq:4da-2}
  \int_0^{\pi} (\sin \omega)^2 \, d \omega
\int_0^\pi \sin \theta\, d \theta \int_0^{2\pi} d \phi\,
\left[ Y^J_{lm} (\omega, \theta, \phi) \right]^*
Y^{J'}_{l'm'} (\omega, \theta, \phi) = \delta_{J,J'} \delta_{l,l'}
\delta_{m,m'}.
\end{equation}


\subsubsection{The parametrization by two $2D$-vectors}
\label{sec:2d-2}

In this section we derive the expression for four-dimensional HSH depending on
the angles of the radius-vector $\vcr$ represented by the two two-dimensional
vectors $\vvr_1$ and $\vvr_2$, 
\begin{equation}
  \label{eq:4d-1}
  \vcr = (\vvr_1,\; \vvr_2), \quad 
\vvr_1 = r_1\, (\cos \phi_1, \, \sin\phi_1), \quad 
\vvr_2 = r_2\, (\cos \phi_2, \, \sin\phi_2),
\end{equation}
where $\phi_1,\phi_2 \in [0, 2\pi)$.
We parametrize the lengths $r_{1,2}$ by the hyperradius $R=\sqrt{r_1^2+r_2^2}$
and the hyperangle $\beta$ as
\begin{equation}
  \label{eq:4d-1a}
  r_1 = R\, \cos\beta, \quad
  r_2 = R\, \sin\beta, \quad \beta \in [0, \pi/2).
\end{equation}
In these coordinates, the integration over the four-dimensional hypersphere is
given by (\ref{eq:gf-11}) which in the above variables reads
\begin{equation}
  \label{eq:4d-1b}
\int d \Omega_4 =  \frac{1}{2} \int_0^{\pi/2} \sin 2 \beta \, d \beta
\int_0^{2\pi} d \phi_1 \int_0^{2\pi} d \phi_2.
\end{equation}
According eqs.~(\ref{eq:gf-12})--(\ref{eq:app-4}) the orthonormal
hyperspherical harmonics take the form
\begin{multline}
  \label{eq:4da-3}
  Y^J_{m_1 m_2} (\hcr) = Y^J_{m_1 m_2} (\beta, \phi_1, \phi_2)
= \frac{1}{\pi} \left[
\frac{(J+1)\,\lambda!\,  (\lambda+m_1+m_2)!}%
{2\, (\lambda+m_1)!\, (\lambda+m_2)!}
\right]^{1/2} e^{i(m_1 \phi_1 + m_2\phi_2)} \\ \times
(\cos \beta)^{m_1} (\sin \beta)^{m_2}
P^{(m_2,\, m_1)}_\lambda (\cos 2\beta),
\end{multline}
where $\lambda= (J-m_1-m_2)/2$ must be an integer number, otherwise 
$Y^J_{m_1 m_2}=0$.
Note also that this definition is valid for $m_1,m_2 \ge 0$.
For negative values of indices the replacement $\phi \to -\phi$ must be used,
e.g.
\begin{equation}
  \label{eq:4da-1b}
 Y^J_{-m_1 m_2} (\beta, \phi_1, \phi_2)
=Y^J_{m_1 m_2} (\beta, -\phi_1, \phi_2).
\end{equation}

By comparing equation (\ref{eq:4da-3}) with the definition of the Wigner
$d$-function (eq.(4.3.4.13) of \cite{Varsh}), one sees that they coincide up
to some constant factor.
Thus, one can choose the four-dimensional HSH to be
\begin{equation}
  \label{eq:4d2-8}
 Y_{j \mu\nu} (\phi_1, \beta, \phi_2)
= e^{i (\mu+\nu) \phi_1 }\, d^j_{\mu \nu} (2 \beta)\, 
  e^{i (\mu-\nu) \phi_2 }
\end{equation}
where $j$ can be both integer and half-integer and the indices
$\mu,\nu=-j,-j+1,\ldots j$.
The connection of the quantum numbers $j\mu\nu$ with $J m_1 m_2$ has the form
\begin{equation}
  \label{eq:4d2-7a}
j=\frac{J}{2}, \quad
 \mu=-\frac{m_1+m_2}{2}, \quad \nu=\frac{m_2-m_1}{2}.
\end{equation}
HSH defined by (\ref{eq:4d2-8}) are orthogonal,
\begin{equation}
  \label{eq:4d2-9}
  \int_{0}^{2\pi} d \phi_1  \int_{0}^{2\pi} d \phi_2
  \int_{0}^{\pi/2} (\sin\beta\, d \beta)\,
  \left[ Y_{j \mu\nu} (\phi_1, \beta, \phi_2) \right]^* \,
  Y_{j' \mu'\nu'} (\phi_1, \beta, \phi_2)
= \frac{8\pi^2}{2j+1} \, \delta_{j,j'} \delta_{\mu,\mu'} \delta_{\nu,\nu'}.
\end{equation}
The scalar product of HSH (\ref{eq:4d2-8}) has the form
\begin{equation}
  \label{eq:4d2-10}
  (Y_j (\hcr) \cdot Y_j (\hcr'))
= \sum_{\mu,\nu=-j}^j {Y_{j \mu \nu}}^* (\hcr) \, Y_{j \mu\nu} (\hcr')
= \chi^j (\cos \omega),
\end{equation}
where $\chi^{j} (\cos \omega)$ is the character of the $O(3)$ group
\cite{Varsh} and 
\begin{equation}
  \label{eq:4d2-10a}
\cos \omega = (\hcr \cdot \hcr')
= \cos\beta \cos\beta' \cos(\phi_1-\phi_1')
+ \sin\beta \sin\beta' \cos(\phi_2-\phi_2').
\end{equation}
where $\omega$ is the angle between $4D$ vectors $\vcr$ and $\vcr'$.
For the sake of completeness we present also the explict epxression for the
character \cite{Varsh} 
\begin{equation}
  \label{4d2-11}
  \chi^j (\cos \omega) = C^1_{2j} (\cos \omega)
= \frac{\sin [(2j+1) \omega]}{\sin \omega}.
\end{equation}
We recall that the rank of HSH $Y_{j \mu\nu}$ defined by (\ref{eq:4d2-8}) is
equal to $2j$.

Thus, we have proved that the four-dimensional hyperspherical harmonics
parametrized by the pair of two-dimensional vectors coincide with the Wigner
$D$-functions which describe the transformation of the three-dimensional
harmonics under the rotation of the coordinate frame \cite{Varsh}.
This fact has many important consequences.
For example, the Clebsch-Gordan coefficients for the four-dimensional HSH can
easily be obtained from the addition theorem for $D$-functions
\cite{avm05:_o4_jpa}.


\subsection{Five-dimensional HSH}
\label{sec:5d}

The five-dimensional space is of less importance from the point of view of
physical applications.
However, it can play a role in the quantum problem of two interacting
particles one of which is moving in the physical space and another one
being restricting to a surface.

Accordingly, it is convenient to parametrize $5D$-vector $\vcr$ as the direct
product of $2D$- and $3D$-vectors,
\begin{equation}
  \label{eq:5d-1}
  \vcr = (\vvr_2, \, \vvr_3)
= R \, ( \cos \beta\, \hr_2, \, \sin \beta\, \hr_3),
\end{equation}
where the unit vectors are
\begin{equation}
  \label{eq:5d-2}
\hr_2 = (\cos \alpha,\, \sin \alpha),\quad
\hr_3 = (\cos \theta,\, \sin \theta \cos \phi, \, 
\sin \theta \sin \phi).
\end{equation}
The corresponding expression for $5D$ HSH can be obtained from
eqs.~(\ref{eq:gf-12}) -- (\ref{eq:app-4}) upon the substitutions $d=5$ and
$\kappa=2$,
\begin{multline}
  \label{eq:5d-3}
  Y^J_{\mu\, lm} (\hcr)
= \left[
\frac{(2J+3)\,\lambda!\,  (J+l+\mu+1)!!}%
{\pi\, 2^{\mu+1}\, (\lambda+\mu)!\, (J+l-\mu+1)!!}
\right]^{1/2}  \\ \times
e^{i \mu \alpha} \, (\cos \beta)^{\mu} (\sin \beta)^{l}
P^{(l+1/2,\, \mu)}_\lambda (\cos 2\beta)\,
Y_{lm} (\theta,\phi), \quad \lambda = \frac{J-l-\mu}{2},
\end{multline}
HSH $Y^J_{\mu\, lm} (\hcr)$ are non-zero only for integer values of
$\lambda$.
The above definition of HSH is valid for $\mu \ge 0$.
In the opposite case the replacement $\alpha \to -\alpha$ must be done in
(\ref{eq:5d-3}) so that
\begin{equation}
  \label{eq:5d-4}
 Y^J_{-\mu\, lm} (\alpha, \beta, \theta,\phi) =  
 Y^J_{\mu\, lm} (-\alpha, \beta, \theta,\phi).
\end{equation}

\subsection{Six-dimensional HSH}
\label{sec:six-dimens-harm}

The dimensionality of the three-body problem after the separation of the
c.m. motion is equal to six.
Furthermore, the dimensionality of the kinetic-energy operator in the
four-body problem can also be reduced to six after the separation of the
collective rotations and c.m. motion.
This explains the importance of $6D$ HSH.

Thus, the natural choice of the arguments of HSH would be the pair of
$3D$ spherical angles plus the hyperangle describing the ratio of $3D$
vectors (sec.~\ref{sec:6d-2d3}).

However, if the planar systems are considered, most useful would be the
parametrization in terms of three planar angles plus two hyperangles
(sec.~\ref{sec:6d-3d2}).


\subsubsection{The parametrization by two three-dimensional vectors}
\label{sec:6d-2d3}

The six-dimensional radius-vector is
$\vcr=R \,(\cos \alpha\, \hr_1,\, \sin\alpha \, \hr_2)$, where
$\hr_{1,2}$ are unit three-dimensional vectors which can be, e.g. Jacobi
vectors of three particles, see fig.~\ref{fig:3bd-jacobi}.

The integration over the solid angle of the six-dimensional hypersphere in
these coordinates has the form
\begin{equation}
  \label{eq:6d1-omega}
  \int d \Omega_6
= \frac{1}{4} \int_0^{\pi/2} (\sin 2 \alpha)^2\, d\alpha \,
\prod_{k=1}^2 \int_0^\pi \sin \theta_k \, d \theta_k
\int_0^{2\pi} d \phi_k,
\end{equation}
where $\theta_k \phi_k$ are the spherical angles of the
unit vector $\hr_k$, $k=1,2$.

The expression for the orthonormal HSH is given by eqs.~(\ref{eq:gf-12}) --
(\ref{eq:app-4}) where $d=6$, $\kappa=3$, so that
\begin{multline}
  \label{eq:6d1-def}
  Y^J_{l_1 m_1\, l_2 m_2} (\alpha, \hr_1, \hr_2) =
\left[
\frac{2^{J+3}\, (J+2)\, \lambda!\, (\lambda+l_1+l_2+1)!}%
{\pi\, (J+l_1-l_2+1)!!\, (J-l_1+l_2+1)!!}
\right]^{1/2} \\ \times
 (\cos \alpha)^{l_1} (\sin \alpha)^{l_2}
P^{(l_2+1/2,\, l_1+1/2)}_\lambda (\cos 2 \alpha)\,
Y_{l_1m_1} (\hr_1) \, Y_{l_2 m_2} (\hr_2).
\end{multline}
Note that HSH are non-zero only for integer values of
$\lambda=(J-l_1-l_2)/2 \ge 0$.

\subsubsection{The parametrization by three two-dimensional vectors}
\label{sec:6d-3d2}

We parametrize the radius-vector as $\vcr = (\vvr_1, \vcr_4)$, where
$\vvr_1=r_1 \, (\cos \phi_1, \, \sin \phi_1)$ and
$\vcr_4$ is $4D$-vector composed of two $2D$-vectors, so that
\begin{equation}
  \label{eq:6d3-1}
\vcr_4 = (\vvr_2, \vvr_3) = (r_2 \cos\phi_2,\,  r_2 \sin\phi_2, \,
r_3 \cos\phi_3,\, r_3 \sin \phi_3).
\end{equation}
It is necessary to specify the parametrization of the lengths $r_1,r_2,r_3$.
We choose it to be
\begin{equation}
  \label{eq:6d3-1b}
  r_1 = R \cos\theta, \quad
  r_2 = R \sin\theta \cos\beta, \quad
  r_3 = R \sin\theta \sin\beta,
\end{equation}
where $\theta,\beta \in [0, \pi/2)$ and
$R^2 = r_1^2+r_2^2+r_3^2$.
The integration over the angles of the six-dimensional hypersphere in the
above coordinates has the form
\begin{equation}
  \label{eq:6d3-1v}
 \int d \Omega_6 = \frac{1}{4}
\int_0^{\pi/2} (\sin \theta)^2 \sin 2\theta\,  d \theta
\int_0^{\pi/2} \sin 2\beta\,  d \beta \,
\prod_{k=1}^3 \int_0^{2 \pi} d \phi_k.
\end{equation}
Using eqs.~(\ref{eq:gf-12})--(\ref{eq:app-4}) where $d=6$, $\kappa=2$ the
orthonormal HSH can be written as
\begin{multline}
  \label{eq:6hsh2-def}
Y^J_{l, m_1 m_2 m_3} (\theta,\beta,\phi_1,\phi_2,\phi_3) =
\left[
\frac{(J+2)\, \lambda!\, (\lambda+m_1+l+1)!}%
{\pi\, (\lambda+m_1)!\, (\lambda+l+1)!}
\right]^{1/2} \\ \times
e^{i m_1 \phi_1} \,
 (\cos \theta)^{m_1} (\sin \theta)^{l}
P^{(l+1,\, m_1)}_\lambda (\cos 2 \theta)\,
Y^l_{m_2 m_3} (\beta,\phi_2,\phi_3).
\end{multline}
These functions are non-zero at non-negative integer values of 
$\lambda=(J-m_1-l)/2$ and $(l-m_2-m_3)/2$.
As a consequence, we have that $(J-m_1-m_2-m_3)/2$ must also be a non-negative
integer number.
In (\ref{eq:6hsh2-def}) the $4D$ HSH $Y^l_{m_2 m_3}(\beta,\phi_2,\phi_3)$ can
be defined by eq.~(\ref{eq:4da-3}) of sec.~\ref{sec:2d-2}.
Again, the above equation (\ref{eq:6hsh2-def}) is valid only form 
$m_1 \ge 0$. 
In the opposite case one has
\begin{equation}
  \label{eq:6d-last}
 Y^J_{l,-m_1 m_2 m_3} (\theta, \beta, \phi_1, \phi_2,\phi_3)
=  Y^J_{l,m_1 m_2 m_3} (\theta, \beta, -\phi_1, \phi_2, \phi_3).
\end{equation}

\section{Conclusion}
\label{sec:conclusion}

In the presented paper the technique of the derivation of the hyperspherical
harmonics with arbitrary arguments has been developed.
This technique does not require the tedious procedure of
the transformation of the Laplace operator to the set of desired variables.
It allows one to obtain the explicit expressions for HSH depending on the
variables which are most suited to the problem under consideration.
For example, in the Helium atom problem the convenient set of variables
comprises the spherical angles and lengths of the position vectors of the
two electrons.
In the calculation of the wave functions of three- and four-electron quantum
dots the set of polar angles plus lengths of position vectors of electrons is
convenient.
In both above problems, HSH may serve as a basis for the expansion of the wave
function in order to transform Schr\"odinger equation to the matrix form.

The main results of the paper are eqs.~(\ref{eq:gf-12})--(\ref{eq:app-4}) and
(\ref{eq:k2-2}), (\ref{eq:k2-4}) of Sec.~\ref{sec:idea} which define
orthogonal and normalized HSH in $d$-dimensional space in terms of products
of lower-dimensional HSH.
As examples of the derived representations, the explicit expressions for
HSH in spaces with dimensions from $2$ to $6$ have been derived in
Sec~\ref{sec:hsh-2-6}.

Expressions for the four-dimensional HSH were presented for the two most
important sets of variables, see Sec.~\ref{sec:3d-1d} and Sec.~\ref{sec:2d-2}.
The importance of four-dimensional HSH stems from the fact that they represent
the wave functions of the hydrogen atom in momentum space \cite{fock35:_o4}.

Six-dimensional HSH depending on the spherical angles of two three-dimensional
vectors and the hyperangle describing their ratio are analyzed in
Sec.~\ref{sec:6d-2d3}.
These HSH are relevant to the quantum three-body problem
\cite{marsh82:_3db_hypersph,matveenko96:_hyperspheroidal}.
Note that 6D HSH given by the expression (\ref{eq:6d1-def}) are not
eigenfunctions of the operator of total angular momentum 
$\mathbf{L}=-i([\vvr_1 \times \nabla_1]+[\vvr_2 \times \nabla_2])$.
The set of HSH being eigenfunction of $\mathbf{L}$ can be constructed by
taking the linear combination of HSH (\ref{eq:6d1-def}) with the conventional
Clebsch-Gordan coefficients \cite{Varsh}.

The expression for the six-dimensional HSH which can be useful in planar
quantum three- and four-body problems is derived in Sec.~\ref{sec:6d-3d2}.
In this case, HSH depend on the polar angles of three co-planar vectors and
the two hyperangles which describe the relative lengths of those vectors.

The application of HSH to $N$-body problems requires the knowledge of
the transformation properties of HSH under the particle exchange.
Such properties depend on the connection of the position vectors of particles
with the hyperangles and in every particular situation must be analyzed
separately.
The procedure of the transformation of HSH under the particle exchange is
often referred to as ``kinematic rotation'' and is discussed e.g. in
\cite{aquilanti86:_ortho_systems,aquilanti:86_hh_3bd}.

We emphasize that the method of zero-length vectors presented in
Sec.~\ref{sec:d-dim-hsh} is quite general and can be used in oder to derive
expressions for \textit{arbitrary} sets of HSH, including non-orthogonal ones.
Once the Cartesian components of the radius vector $\vvr$ are parametrized in
terms of the hyperradius and hyperangles, $J$-th rank HSH will be given by the
coeffcients in the expansion of the function $(\vvr \cdot \va)^J$ where $\va$
is an arbitrary constant zero-length vector, $(\va \cdot \va)=0$.
Probably, this gives the most simple approach to the calculation of HSH.

Finally, we note that the method proposed in this paper can also be applied to
the problem of the calculation of Clebsch-Gordan coefficients in
many-dimensional space.
These coefficients allow one to evaluate many-dimensional integrals
involving the products of HSH.
Clebsch-Gordan coefficients are also necessary for the derivation of the
multipole expansions of functions depending on several vector arguments.
Examples of such multipole expansions of functions depending on 
$|\vcr - \vcr'|$ in three- and four-dimensional space may be found in
\cite{avm05:_o4_jpa,manakov02:_multip}.


\section{Acknowledgments}
\label{sec:acknowledgments}

This work has been supported in part by the joint BRHE program of CRDF and
Russian Ministry of Education, by the grant from ``Dynasty foundation''
and by the grant MK-862.2007.2 from President of Russia.

\appendix

\section{The explicit form of $h^J_{ll'}$ and $g^J_l$}
\label{sec:app-hj}

In this section we calculate the functions $h^J_{ll'} (\theta_\kappa)$ defined
by eq.~(\ref{eq:gf-8}) of the main text.

From the properties of the $B$-coefficients (\ref{eq:gf-5a}) it follows that
$h^J_{ll'} (\theta_\kappa)=0$ for the combination $(J-l-l')=2\lambda$ being an
\textit{odd} number.
Thus, $\lambda$ must always be an integer.

Substituting eq.~(\ref{eq:gf-5a}) and the similar identity for 
$B^{(d-\kappa)}_{(J-q)l'}$ into eq.~(\ref{eq:gf-8}), after some simple
transformations, we obtain
\begin{multline}
  \label{eq:app-1}
 h^J_{ll'} (\theta_\kappa) = i^{l'} 2^{-J}
\sum_{n=0}^\lambda (-1)^{\lambda+n}\,
\frac{(l+\kappa/2-1)\,\Gamma(\kappa/2-1) }%
{n!\, (\lambda-n)!\, \Gamma(n+l+\kappa/2)} \\ \times
\frac{(l'+(d-\kappa)/2-1)\,\Gamma((d-\kappa)/2-1)}%
{\Gamma(\lambda-n+l'+(d-\kappa)/2)} 
(\cos \theta_\kappa)^{l+2n}\, (\sin \theta_{\kappa})^{l'+2\lambda-2n}
\end{multline}
Here, the summation leads to the Jacobi polynomial \cite{Bateman-II},
\begin{equation}
  \label{eq:app-2}
 h^J_{ll'} (\theta_\kappa)
=  A^{(d,\kappa)}_{\lambda l l'}\,
(\cos \theta_\kappa)^{l}\, (\sin \theta_{\kappa})^{l'}\,
P_{\lambda}^{(l'-1 + \frac{d-\kappa}{2}, \,
 l -1 + \frac{\kappa}{2})} (\cos 2 \theta_\kappa).
\end{equation}
where the coefficient $ A^{(d,\kappa)}_{\lambda l l'}$ is
\begin{equation}
  \label{eq:app-2a}
  A^{(d,\kappa)}_{\lambda l l'}
= \frac{i^{l'}\, (l+\kappa/2-1) \, ( l'+ (d-\kappa)/2-1 ) \,
\Gamma ( \kappa/2-1 )\, \Gamma ( (d-\kappa)/2-1 ) }%
{2^J\, \Gamma(\lambda+l'+ (d-\kappa)/2) \,
 \Gamma(\lambda+l + \kappa/2) }.
\end{equation}

The explicit expression for the function $g^J_l$ defined in (\ref{eq:k1-6})
is 
\begin{equation}
  \label{eq:app-g}
  g^J_l (\theta) = i^l\, 2^{-J} J!\, \left( l+ \frac{d-3}{2} \right)\,
\Gamma \left( \frac{d-3}{2} \right)
\sum_{n=0} \frac{(-1)^n \, (\sin \theta)^{l+2n} \, (2\cos \theta)^{J-l-2n}}%
{n!\, (J-l-2n)!\, \Gamma(l+n+d/2)}.
\end{equation}
The sum over $n$ evaluates to the Gauss hypergeometric function which is
equivalent to the Gegenbauer polynomial \cite{Bateman-II},
\begin{equation}
  \label{eq:app-g-2}
  g^J_l (\theta) =   A^{(d)}_{J l}\,
(\sin \theta)^{l}\,
C_{J-l}^{l + \frac{d-1}{2}} (\cos \theta),
\end{equation}
where 
\begin{equation}
  \label{eq:app-g-3}
  A^{(d)}_{J l}
=  i^l\, 2^{d+l-J-2} J!\, \left( l+ \frac{d-3}{2} \right)\,
\Gamma \left( \frac{d-3}{2} \right)
\frac{\Gamma(l+ (d-1)/2)}{\Gamma(J+l+d-1)}.
\end{equation}


\end{document}